\documentstyle[agupp,epsfig]{article}
\lefthead{GREATBATCH AND PETERSON}
\def
\righthead{INTERDECADAL THERMOHALINE ADJUSTMENT}
\received{June 26, 1995}
\revised{April 17, 1996}
\accepted{May 14, 1996}
\journalid{JGRA}{}
\articleid{}{}
\paperid{96JC01531}
\ccc{0148-0227/96/96JC-01531\$05.00}

\authoraddr{R.J. Greatbatch and K.A. Peterson, Department of
Physics and Physical Oceanography, Memorial University of
Newfoundland, St. John's, NF, Canada A1B 3X7 (email:
rgreat@crosby.physics.mun.ca and drew@crosby.physics.mun.ca).}

\slugcomment{
To appear in the {\it Journal of Geophysical Research (Oceans)}, 1996.
}

\newcommand{\BA}{\begin{eqnarray}}
\newcommand{\EA}{\end{eqnarray}}

\begin{document}

\title{\bf Interdecadal variability and oceanic thermohaline
adjustment}

\author{
Richard J. Greatbatch
and K. Andrew Peterson \altaffilmark{1} \altaffilmark{2}
}
\affil{Department of Physics and Physical Oceanography,
Memorial University of Newfoundland, St. John's, Newfoundland,
Canada}
\altaffiltext{1}
{
Address: R.J. Greatbatch and K.A. Peterson, Department of
Physics and Physical Oceanography, Memorial University of
Newfoundland, St. John's, NF, Canada A1B 3X7 (email:
rgreat@crosby.physics.mun.ca and drew@crosby.physics.mun.ca).
}
\altaffiltext{2}
{
This paper is also available from
ftp://crosby.physics.mun.ca/pub/drew/papers/gp1.ps.gz
}

\begin{abstract}
Changes in the strength of the thermohaline overturning
circulation are associated,
by geostrophy, with changes in the east-west pressure difference
across an ocean basin.
The tropical-polar density contrast and the east-west pressure
difference are
connected by an adjustment process. In flat-bottomed ocean models
the adjustment is associated with viscous, baroclinic Kelvin wave
propagation. Weak-high latitude stratification leads to the
adjustment having an interdecadal timescale.
We reexamine model interdecadal oscillations in the context of
the adjustment process, for both
constant flux and mixed surface boundary conditions.
Under constant surface flux,
interdecadal oscillations are associated with the passage of a
viscous Kelvin wave around
the model domain. We carry out experiments suppressing wave
propagation along each
of the model boundaries. Suppressing wave propagation
along either the tropical or eastern boundary
does not eliminate the oscillation, but increases both its period
and amplitude.
Suppressing wave propagation along either the polar or the
western
boundary eliminates the oscillation. Our results suggest the
oscillations can be
self-sustained by perturbations to
the western boundary current arising from the southward boundary
wave propagation.
Mixed boundary condition oscillations
are characterized by the eastward, cross-basin movement of
salinity-dominated density anomalies, and the westward return of
these anomalies
along the northern boundary. We suggest the latter is associated
with
viscous Kelvin wave propagation.
Under both types of boundary conditions, the strength of the
thermohaline
overturning and the tropical-polar density contrast vary out of
phase.
We show how the
phase relationship is related to the boundary wave propagation.
Box models and
zonally averaged models assume that the east-west and north-south
pressure gradients
vary in phase and are proportional to one another. We suggest
this assumption is valid only on timescales long compared to the
adjustment timescale. The importance of boundary regions
indicates
an urgent need to examine the robustness of interdecadal
variability
in models as the resolution is increased, and as the
representation of the coastal,
shelf/slope wave guide is improved.
\end{abstract}

\section{\bf 1. Introduction}

Interdecadal variability is a fundamental
feature of the climate system (see \markcite{{\it Weaver and
Hughes} [1992]}
for a review, and the
recent papers by \markcite{{\it Deser and Blackmon} [1993]},
\markcite{{\it Kushnir} [1994]}, and \markcite{{\it Latif and
Barnett} [1994]}).
Interdecadal variability has also been found in coarse resolution
(i.e., non-eddy resolving) ocean models.
\markcite{{\it Marotzke}
[1990]} and
\markcite{{\it Weaver and Sarachik} [1991]} were the first to
find such variability.
These authors
used mixed boundary conditions to represent the ocean/atmosphere
interaction;
that is, a strong (tens of days timescale) restoring boundary
condition on the
surface temperature and a constant flux boundary condition on the
surface salinity.
Interdecadal variability has also been found in models run under
constant surface
buoyancy flux [\markcite{{\it Huang and Chou}, 1994};
\markcite{{\it Greatbatch and Zhang}, 1995};
\markcite{{\it Cai et al.}, 1995};
\markcite{{\it Chen and Ghil}, 1995}] and in fully coupled
ocean/atmosphere models [\markcite{{\it Delworth et al.}, 1993};
\markcite{{\it Latif and Barnett}, 1994}].
In addition, \markcite{{\it Weaver et al.}\ [1994]} have
described
an interdecadal
oscillation in a coarse resolution North Atlantic model.
\markcite{{\it Greatbatch and Zhang} [1995]} have noted the
strong
similarity between their oscillation under constant heat flux,
and the
oscillation found in the Geophysical Fluid Dynamics Laboratory
(GFDL) coupled model [\markcite{{\it Delworth et al.}, 1993}].
\markcite{{\it Griffies and Tziperman} [1995]} have offered an
alternative
interpretation of the Delworth et al.\ oscillation using a
four-box model run under
mixed surface boundary conditions.

Central to interdecadal variability in ocean models are
interdecadal
fluctuations in the thermohaline circulation. The thermohaline
circulation
is driven by the formation of dense water in high latitudes in
response to
intense surface cooling. The new dense water spreads equatorward
and is replaced at the
surface by warm, salty water from lower latitudes. Consider a
flat-bottomed
ocean basin with both eastern and western boundaries.
Assuming the north/south flow to be in geostrophic balance, then
the northward
transport per unit depth through a line of latitude $\phi$ and at
depth {\it z} is given by
\begin{equation}
V(\phi, z) = {{({p_E} - {p_W})} \over {f{\rho}_0}}
\end{equation}
where $p_E$ and $p_W$ are the pressure on the eastern and western
boundaries,
${\rho}_0$ is a representive density for seawater, and
$f$ is the Coriolis parameter. It follows that to understand
fluctuations in the thermohaline overturning, it is necessary to
understand
what controls the east-west pressure difference across an ocean
basin.
Undoubtedly, $p_E$ and $p_W$ will be strongly influenced by both
coastal trapped
wave propagation and advective processes in the boundary region
(e.g., the Deep Western
Boundary Undercurrent in the case of the North Atlantic).
Equation (1) points to a
fundamental issue in thermohaline ocean circulation theory,
namely, that of how
a north/south density gradient, established by high-latitude deep
water
formation, can set up the east/west pressure gradient necessary
to maintain the
geostrophic balance of the north/south flowing thermohaline
circulation.
\markcite{{\it Zhang et al.}\ [1992]} offer an explanation in
terms of a
``primary/secondary circulation"
argument. The ``primary circulation" is in thermal wind balance
with the north/south
density gradient, with eastward flow at the surface, and westward
return flow beneath.
The meridional boundaries block the primary circulation, leading
to downwelling on the
eastern boundary, and upwelling on the western boundary. This
sets up east/west
pressure gradients that then drive the north/south flow of the
thermohaline circulation.

\markcite{{\it Zhang et al.}\ [1992]} did not discuss the details
of the interaction
between
the primary circulation and the boundaries. In particular, they
did not consider
the role played by the coastal wave guide. In an early paper,
\markcite{{\it Davey} [1983]} described the
spin-up of a two-level model driven by a restoring surface
boundary condition on
temperature. The restoring temperature varied in the north/south
direction.
Davey noted that including meridional boundaries had a major
impact on the spin-up,
leading to an adjustment process involving coastal Kelvin waves
and Rossby waves.
\markcite{{\it Wajsowicz and Gill} [1986]} then went on to
consider the spin-down of
an ocean model initialized with a specified density field that
varied only in the
north-south direction. For their choice of initial density field,
the first stage of the spin-down is accomplished by
baroclinic, coastal Kelvin waves and takes place during the first
few months of
the model adjustment. Wajsowicz and Gill noted that these waves
are severely corrupted
by the coarse
resolution and the large horizontal eddy viscosity commonly used
in models
[\markcite{{\it Hsieh et al.}, 1983}]. The second stage of the
spin-down
takes place on a decadal timescale and is associated with the
propagation of
long, baroclinic Rossby waves [\markcite{{\it Wajsowicz}, 1986}].

Recently, \markcite{{\it Winton} [1996]} has revisited the
problem
considered by \markcite{{\it Wajsowicz and Gill} [1986]}.
Winton noted the importance of viscous boundary waves
in models that exhibit interdecadal variability. The viscous
boundary waves
identified by Winton are the low-frequency form of the viscous,
baroclinic Kelvin
waves noted by  \markcite{{\it Wajsowicz and Gill} [1986]}. At
first sight,
it may seem surprising that
a Kelvin wave propagating along the boundary of a model domain
could be important
on an interdecadal timescale. If we consider a model
extending from the equator to $60^o$N, and of $60^o$ width in
longitude, then a Kelvin
wave propagating at 1 m s$^{-1}$ (a typical speed for the first
baroclinic mode)
would take 0.7 years to travel the length of the
boundary. Obviously, this timescale is considerably shorter than
interdecadal.
The key to the discrepancy rests with the model stratification,
which is usually
weak or nonexistent in the high latitudes owing to the presence
of deep,
convective mixing [\markcite{{\it Winton}, 1996}]. The weak
stratification impedes
wave propagation and leads
to the emergence of an interdecadal timescale.
\markcite{{\it Wajsowicz and Gill} [1986]} did not notice this
effect because
the initial density stratification used in their spin-down
experiment was
not weak enough at high latitudes. The weak stratification
can also lead to highly nonlinear behavior in the high latitudes
and may be
responsible for much of the sensitivity found in models [e.g.,
\markcite{{\it Huang and Chou}, 1994}].

Once it is appreciated that Kelvin wave adjustment in models has
an interdecadal
timescale,
a question immediately arises as to the robustness of
interdecadal variability in
coarse-resolution ocean models.
The inability to resolve the internal radius of deformation
(especially in high
latitudes where this is measurable in kilometers),
the use of large eddy viscosity and diffusivity parameters, the
inadequate representation of the ocean bottom topography, and the
inadequate
representation of boundary currents such as the Deep Western
Boundary Undercurrent all
preclude coarse-resolution models from resolving boundary
processes as they occur in
nature.
A similar problem is the inability of models to adequately
resolve the process by which
North Atlantic Deep Water spills over the
Greenland/Iceland/Scotland ridge and makes
its way into the North Atlantic and the global ocean circulation
[e.g., \markcite{{\it Killworth}, 1992]}.

Recently, \markcite{{\it D{\"o}scher et al.}\ [1994]}
have described the adjustment process in high-resolution
eddy-resolving and
non-eddy-resolving models of the North Atlantic. These authors
considered the model
response to a change in the
temperature and salinity boundary condition applied along the
northern boundary.
They showed that the adjustment involves both wave and advective
processes along the
coastal and equatorial wave guides. They also showed that the
adjustment process
depends on model resolution. In a related study,
\markcite{{\it Gerdes and K{\"o}berle} [1995]}
have examined the response of a $1^o$ x $1^o$ model of the North
Atlantic to a change
in the surface temperature and salinity in the Denmark Strait
region.
Once again, the model shows a response in terms of both
coastal trapped wave propagation and advective processes. In both
these studies, the
coastal trapped wave response is rapid, the waves excited in the
northern North
Atlantic reaching the equator on a timescale of weeks to months.
This is followed by
a much slower advective adjustment on a timescale of tens of
years.
These higher resolution studies suggest that the internal
adjustment
timescale of the ocean is indeed decadal, but they also suggest
that in nature,
advective processes may be more important than wave processes. On
the other hand,
neither of the problems studied in these papers involves wave
propagation along a weakly
stratified, high-latitude boundary.

A question fundamentally related to the adjustment process is
that of the relationship
between the strength of the thermohaline overturning circulation
and the north/south
density gradient. In box models [e.g.,
\markcite{{\it Stommel}, 1961; {\it Griffies and Tziperman},
1995}]
it is assumed that the
former is directly proportional to the latter, with no phase lag.
There
is evidence, in models at least, that this relationship does not
hold on
interdecadal timescales. This is illustrated by Figure 1, taken
from
\markcite{{\it Greatbatch and Zhang} [1995]}. The figure shows
anomalies (that is,
difference from the mean)
in the overturning stream function (left panels) and the zonally
averaged temperature
(right panels). The salinity is uniform, so that warmer/colder
temperatures
are associated with lighter/denser water. At the time in Figure
1b, the high latitudes are warmer,
and therefore less dense, than in the mean, indicating a reduced
density contrast
between the equatorial and polar regions; yet, at this time, the
overturning circulation
in the basin reaches its maximum value.
\markcite{{\it Zhang et al.}\ [1995]} describe a decadal
oscillation
in a coupled thermodynamic sea-ice/ocean circulation model that
also shows some
interesting phase relationships. In this case, the maximum in the
overturning circulation
is associated with the minimum in the convective overturning
activity, and the minimum
in the high-latitude surface heat loss.

\begin{figure}[tbp]
\figurewidth{20pc}
\figurenum{1}
\begin{center}
\mbox{\epsfig{figure=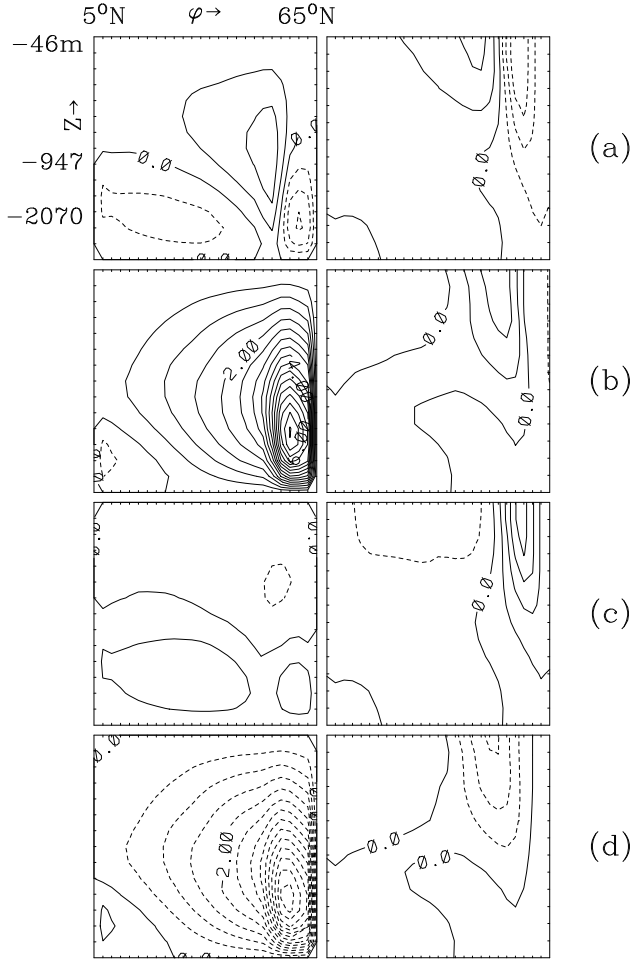,width=8.4cm}} 
\end{center}
\caption{Anomalies (that is, difference from the mean) of the
overturning stream function (left column) and the zonally
averaged temperature (right column) taken from \markcite{{\it
Greatbatch and Zhang} [1995]}.
Figures 1a, 1b, 1c, and 1d are each 12.5 years apart, with the
meridional overturning being a maximum at Figure 1b and a minimum
at Figure 1d. The contour
intervals are 0.5 Sv and 0.5$^o$C, respectively. Negative
anomalies are shown
using dashed contours. The vertical scale is expanded in the
upper part of the water
column.}
\end{figure}

In this paper, we reexamine interdecadal variability
in ocean-only models. We begin with the thermal-only, constant
surface
flux oscillation
of \markcite{{\it Greatbatch and Zhang} [1995]}. Since there is
no variation
in the surface forcing, oscillations under fixed surface flux
highlight the role of the
internal adjustment process. We confirm
\markcite{{\it Winton}'s [1996]} conclusion that boundary waves
play an important
role in these oscillations. We explore the
influence of wave propagation along each of the model boundaries
and suggest a mechanism
by which the oscillations can be self-sustained by perturbations
to the western boundary current. We also discuss the phase
relationship between
the strength of the thermohaline overturning and the
tropical-polar density contrast,
and comment on the validity of parameterizations which assume
these two quantities
to be in phase and
proportional. We also show that ``zonal redistribution," as
discussed
by \markcite{{\it Cai et al.}\ [1995]}
is not always necessary for interdecadal oscillations
to occur. In particular, we show results from a model run in
which a self-sustained
interdecadal oscillation occurs when the constant flux forcing is
the same as the flux
diagnosed from a spin-up experiment, all model parameters, model
geometry, etc., being
the same as in the spin-up. Finally, we reexamine decadal
oscillations under
mixed boundary conditions
[e.g., \markcite{{\it Weaver and Sarachik}, 1991]}. The situation
is
now more complex because of the variable surface heat flux
associated
with the restoring boundary condition on the surface temperature.
The role of
boundary wave propagation along the model boundaries is
demonstrated, and once again,
we find that the strength of the thermohaline overturning and the
tropical-polar density
contrast vary out of phase with each other.

The structure of this paper is as follows. In section 2 we
describe
the model experiments.
Section 3 gives the model results from cases run with thermal
forcing only (uniform
salinity). This section is divided into several parts: a
small-amplitude oscillation,
a large-amplitude oscillation, and
the role of propagation along each of the model boundaries, where
we also discuss
how the oscillations are maintained.
Section 4 discusses mixed boundary conditions, and section 5
provides
a summary and discussion.

\section{\bf 2. Model Description}

\begin{table}[t]
\tablenum{1}
\tablewidth{8.4 cm}
\caption{The Depths of the Center of Each Model Level}
\begin{center}
\begin{tabular*}{6.0cm}{c@{\extracolsep{\fill}}r}
\hline
\mbox{}\vspace{-2mm} \\
\vspace{1mm}
{Level} &
{Depth, m}
\\ \hline
\vspace{-2mm} \nl
1& 23.0\nl
2& 75.0\nl
3& 140.5\nl
4& 223.0\nl
5& 327.0\nl
6& 458.0\nl
7& 623.0\nl
8& 831.0\nl
9& 1093.0\nl
10& 1423.0\nl
11& 1838.5\nl
12& 2362.0\nl
13& 2990.5\nl
\vspace{1mm}
14& 3663.5
\\ \hline
\end{tabular*}
\end{center}
\end{table}

We use a primitive equation, spherical coordinate model
\markcite{{\it Greatbatch et al.}\ [1995]}, very similiar to the
Bryan-Cox-Semtner model
[\markcite{{\it Bryan}, 1969}; \markcite{{\it Cox}, 1984};
\markcite{{\it Semtner}, 1974}]. A realistic equation of state is
used,
like that of
\markcite{{\it Bryan and Cox} [1972]}. The model domain is a
flat-bottomed (4000 m depth)
basin, extending from $5^o$N\ to $65^o$N, with a longitudinal
extent of $60^o$.
All the model experiments use the same 14 levels in the vertical
(the levels are given in
Table 1), and have $2.4^o$ x $2.4^o$ horizontal resolution.
In addition, all experiments use uniform values of
$10^{-3}$ m$^2$ s$^{-1}$ and $10^{-4}$ m$^2$ s$^{-1}$ for the
vertical eddy viscosity and
diffusivity, respectively, and are run with the wind forcing set
to zero. Deep
convection is parameterized as in the work by \markcite{{\it Cox}
[1984]}, by using a large value
($10^5$ m$^2$ s$^{-1}$) for the vertical diffusivity whenever a
hydrostatically
unstable density profile is generated.
The model experiments are listed in Table 2 and differ in their
surface forcing and
model parameters.

\begin{planotable}{rrrrrrrr}
\tablenum{2}
\tablewidth{16.8 cm}
\tablecaption{The Model Experiments}
\tablenum{2}
\tablehead{
     \colhead{Experiment}&
     \colhead{Horizontal}&
     \colhead{Horizontal}&
     \colhead{Grid}&
     \colhead{Surface}&
     \colhead{Period,}&
     \colhead{Maximum}&
     \colhead{Mean}\\
     \colhead{}&
     \colhead{Diffusivity,}&
     \colhead{Viscosity,}&
     \colhead{Size,}&
     \colhead{Flux}&
     \colhead{}&
     \colhead{Amplitude,}&
     \colhead{Maximum,}\\
     \colhead{}&
     \colhead{m$^2$ s$^{-1}$}&
     \colhead{10$^{3}$ m$^2$ s$^{-1}$}&
     \colhead{deg}&
     \colhead{}&
     \colhead{years}&
     \colhead{Sv}&
     \colhead{Sv}}
\tablenotetext{\null}{D, diagnosed flux; Z, zonal average of the
diagnosed flux; R, restoring boundary condition. When two letters
are used, the first refers to the heat flux, the second to the
freshwater flux.}

\tablenotetext{a}{Wave propagation suppressed on the southern
boundary.}

\tablenotetext{b}{Wave propagation suppressed on the eastern
boundary.}

\tablenotetext{c}{Wave propagation suppressed on the northern
boundary.}

\tablenotetext{d}{Wave propagation suppressed on the western
boundary.}

\startdata
   A0& 2000& 100.0& 2.4& D&      0.0&    0.0& 18.5 \nl
   A1& 2000& 100.0& 2.4& Z&     33.9&    2.6& 15.0 \nl

  B0& 1000& 100.0& 2.4&  D&    30.5&    7.7& 17.3 \nl
 B0a\tablenotemark{a}
    & 1000& 100.0& 2.4&  D&    37.2&   11.5& 19.9 \nl
B0b\tablenotemark{b}
    & 1000& 100.0& 2.4&  D&    36.2&   10.3& 20.6 \nl
 B0c\tablenotemark{c}
    & 1000& 100.0& 2.4&  D&     0.0&    0.0& 18.3 \nl
 B0d\tablenotemark{d}
    & 1000& 100.0& 2.4&  D&     0.0&    0.0& 18.2 \nl
 C1& 1000& 100.0& 2.4& RD&   26-17& 0.4-1.6& 2.3-6.0 \nl
\end{planotable}

We shall focus on two kinds of interdecadal variability found in
ocean-only models, namely, the oscillations found under constant
heat flux
by \markcite{{\it Greatbatch and Zhang} [1995]} and oscillations
found under mixed
boundary conditions by \markcite{{\it Weaver and Sarachik}
[1991]}. It follows that the
model experiments are divided into two groups, as described
below.

\subsection{\bf 2.1. Temperature-Only Cases}

In the temperature-only set of experiments (labeled A and B in
Table 2)
the model is run using a uniform value for the
salinity of $35^o/_{oo}$. There is no freshwater flux forcing.

\begin{figure}[tbp]
\figurenum{2}
\figurewidth{20pc}
\begin{center}
\mbox{\epsfig{figure=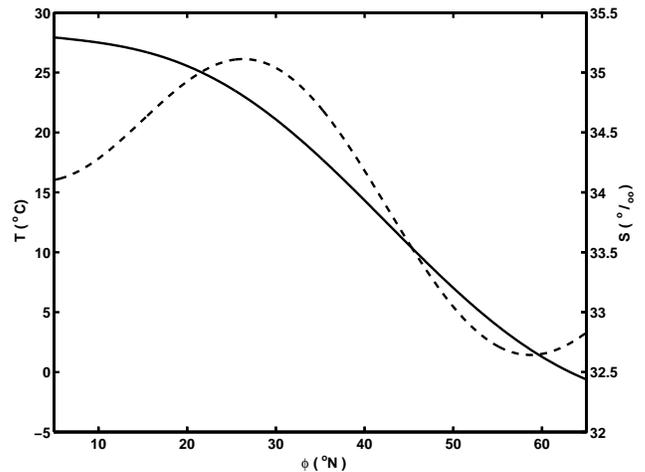,width=8.4cm}} 
\end{center}
\caption{Zonal restoring temperature and salinity as a function
of latitude.  Solid line is temperature; dashed line is
salinity.}
\end{figure}

First a restoring spin-up experiment is carried out, restoring
the temperature
in the top level of the model to a specified, zonally uniform
restoring temperature
given by
\BA
T_a(\phi) &=& 30.0 - 33.0/80.0 \times \phi \nonumber\\
        &+& 4.0 \times \sin{( (2\pi/75.0) (\phi - 5.0) )},
\EA
where $\phi$\ is latitude in degrees (see Figure 2).
The restoring timescale is 30 days.
To increase the speed with which the model is brought to
equilibrium,
the acceleration techniques of \markcite{{\it Bryan} [1984]} are
used.
The acceleration is turned off for at least the last 500 years of
the
spin-up, and in all subsequent experiments. The exception
is a mild distortion on the momentum equations (the local
time derivative terms in the momentum equations are multiplied by
a factor of 10).
The latter allows
the use of a longer time step than would otherwise be possible,
and has no effect
on the model results.

Following the spin-up, the surface heat flux into the ocean (in
W m$^{-2}$)
is diagnosed by calculating the average of $Q_T$, defined below,
over the
last 50 years of the spin-up. $Q_T$ is given by
\begin{equation}
Q_T = {\rho_0}{c_p}{\gamma_T} (T_a - T_1) \delta z ,
\end{equation}
\noindent
where $T_a$\ is the restoring temperature given by equation (2),
$T_1$\ is the temperature in the top level of the model,
$\gamma_T$\ is the restoring time constant
(1/$\gamma_T$ = 30 days),
$\delta z$ is the thickness of the top layer ($46$ m, and
${\rho_0}{c_p}$
is density multiplied by specific heat at constant pressure.
The diagnosed heat flux
is then used to drive the model in subsequent experiments.

\subsection{\bf 2.2. Temperature and Salinity Cases}

In this set of experiments (labeled C in Table 2),
the model is spun-up to equilibrium using
restoring conditions applied to both the surface temperature
and surface salinity.  The restoring temperature is as before.
The surface
salinity is restored to a zonally uniform field (see Figure 2)
given (in $^o/_{oo}$) by
\BA
S_a(\phi) &=& 35.0 - 1.32/50.0 \times \phi \nonumber\\
        &+& 0.84 \times \sin{( (2\pi/55.0) (\phi - 15.0) )}.
\EA
A restoring timescale of 30 days is used, as for temperature.
Upon reaching equilibrium, the corresponding surface heat and
``virtual" salt fluxes are
calculated (note that we use a virtual salt flux to drive the
model, rather than
the more realistic freshwater flux, as described by
\markcite{{\it Huang} [1993]}).
In analogy to equation (3), the surface salt flux is diagnosed
from the last 50 years
of the spin-up using
\begin{equation}
Q_S = \gamma_S (S_a - S_1) \delta z ,
\end{equation}
where $S_a$\ and $S_1$\ are the restoring and top level model
salinities
respectively,
$\gamma_S$\ is the restoring time constant
(1/$\gamma_S$ = 1/$\gamma_T$ = 30 days),
and $Q_S$\ is given in $^o/_{oo}$~m~s$^{-1}$.

\subsection{\bf 2.3. Model Diagnostics}

The most useful diagnostic is the baroclinic pressure defined by
\begin{equation}
P = p - {1 \over H}{{\int}_{-H}^0}pdz
\end{equation}
where
\begin{equation}
p = g{{\int}_{z}^0}{\lbrace}{\rho} -
{\overline{\rho}}(z^{\prime}){\rbrace}{dz^{\prime}}
\end{equation}
${\overline{\rho}}(z)$ is a reference density field that depends
only on the vertical
coordinate, $z$, and is usually taken to be the density averaged
horizontally over the
model domain and averaged in time over several oscillation
periods.
Since $P$ is the baroclinic pressure, it has zero vertical
average.
It should be noted that because we do not include wind forcing,
the barotropic flow in our model is very weak. As a consequence,
the horizontal
gradients of $P$ are a very good approximation to the horizontal
gradients
of the total pressure.

\begin{figure*}
\figurenum{3}
\figurewidth{35pc}
\begin{center}
\mbox{\epsfig{figure=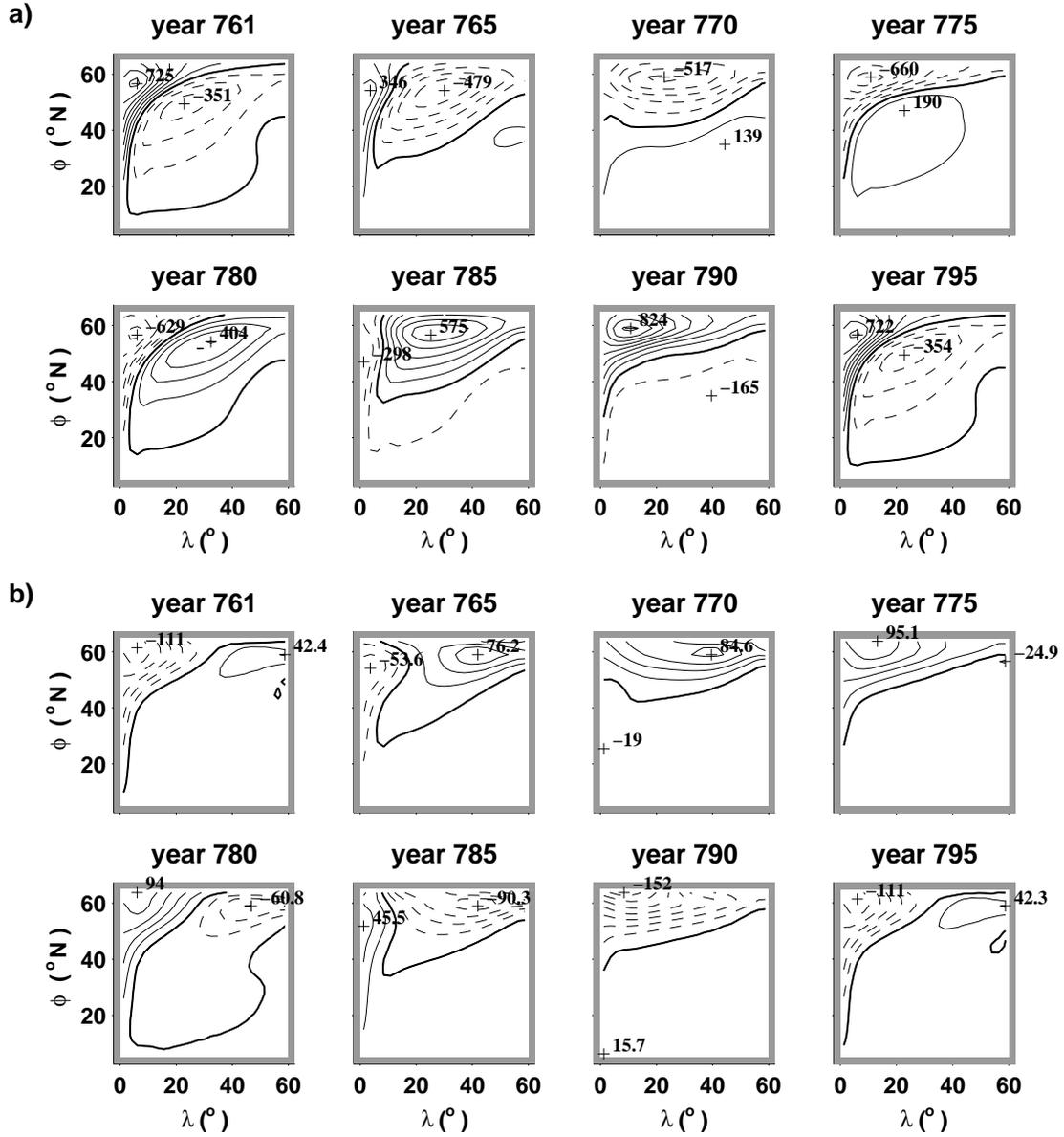,width=14.7cm}} 
\end{center}
\caption{Experiment A1. Snapshots of baroclinic pressure
anomalies
(that is, difference from the mean averaged between years 612 and
984)
for (a) level 1 (0 m to 46 m deep) with contour interval of 100
Pa and (b) level 12 (2070 m to 2654 m deep) with contour interval
of 20 Pa.
Negative values are shown using dashed contours.
The maximum and minimum anomalies are displayed on contours (in
Pa) and identified with a cross.}
\end{figure*}

\section{\bf 3. Oscillations Under Constant Flux}

\subsection{\bf 3.1. A Small-Amplitude Oscillation}

We begin with Experiment A0 in Table 2. The model is first
spun-up,
as described in section 2. When the surface boundary condition is
replaced by the
diagnosed heat flux, and the model initialized with the end of
the
spin-up (all model parameters the same as in the spin-up),
no oscillations are found; in particular, the model state remains
unchanged. As discussed by \markcite{{\it Cai et al}. [1995]},
oscillations can be induced by
zonally redistributing the diagnosed flux (that is, replacing the
diagnosed flux by a linear combination of the diagnosed flux and
its zonal average).
For simplicity, we restrict attention to the case in which the
model is driven by
the zonal average of the diagnosed flux (Experiment A1).
Initializing
with the end state of the spin-up, the model develops a steady
oscillation with
period 33.9 years. It should be noted that although we show
results from model
experiments that exhibit self-sustained oscillations, damped
oscillations are also
found if different model parameters and/or model resolution are
used. Model results
are particularly sensitive to the value of the horizontal
diffusivity, an example
of which is given in section 3.2 (examples of damped oscillations
can also be found in the work by \markcite{{\it Cai et al}.
[1995]}).

Figure 3 shows snapshots of the baroclinic pressure (defined by
equation (6))
at the surface (Figure 3a) and at level 12 (Figure 3b; 2362-m
depth) from
Experiment A1.
Anomalies in baroclinic pressure are plotted (that is, with the
time average over
many oscillations removed). Comparing the two depths, we see that
the pressure
anomalies are generally in phase but of opposite sign.
Examination
of the anomalies at other depths confirms the impression that the
sign of the anomalies
changes only once over the depth of the  water column, indicative
of the first
baroclinic normal mode [\markcite{{\it Gill}, 1982}] (it should
be noted
that the depth of the zero crossing varies spatially, as one
might expect, given the
large spatial variations in the mean density field). The
anomalous flow field associated
with the anomalous pressure fields can be easily estimated from
geostrophy. For example,
at year 761, anomalous pressure is high at the surface on the
western side of the
model basin, and relatively low on the eastern side, implying
anomalous southward flow
at the surface and northward flow below (recall equation (1)).
It is therefore not surprising to find
that year 761 coincides with a minimum in the strength of the
meridional overturning,
whereas year 780, when the pressure pattern is reversed, is
associated with a maximum
in the overturning.

Associated with the anomalies in the strength of the overturning
are strong gradients in anomalous baroclinic pressure along the
model boundaries.
For example, at year 761 (780), when the overturning strength is
a minimum (maximum),
there is an anomalous east-west pressure gradient along the
northern boundary, with
surface geostrophic flow away from (toward) the boundary.
Geostrophic flow normal
to the model boundary is associated with propagation along the
boundary, as has
been discussed by  \markcite{{\it Winton} [1996]}. For example,
when surface flow
is toward the northern boundary, as at year 780, heat is advected
toward the boundary.
The convergence at the boundary warms the water column, raising
the baroclinic pressure
at the surface, and lowering it below. Since the flow is
associated with high
baroclinic pressure to the east, and low baroclinic pressure to
the west, the result is
a westward propagation of the pressure gradient associated with
the flow.
\begin{figure}[tbp]
\figurenum{4}
\figurewidth{20pc}
\begin{center}
\mbox{\epsfig{figure=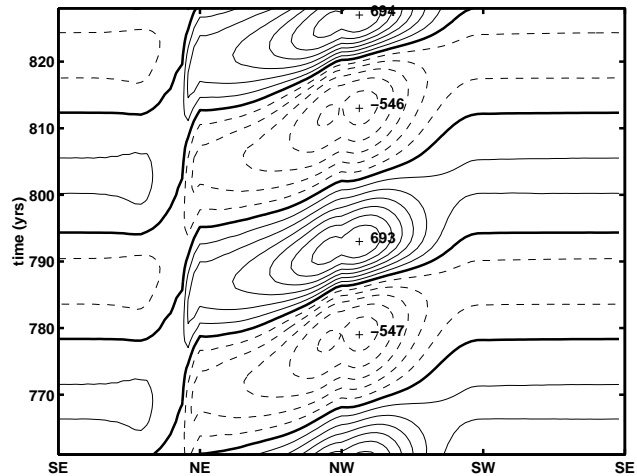,width=8.4cm}} 
\end{center}
\caption{Experiment A1. Contours of baroclinic pressure anomaly
at level 1 (surface) as a function of
distance along the model boundary and time.  Distance is measured
in a counterclockwise direction from the southeast corner (marked
SE). The contour interval is 100 Pa.
Negative values are shown using dashed contours.
The maximum and minimum values are displayed (in Pa) for each
oscillation and identified with a cross.}
\end{figure}
Propagation is clearly evident in Figure 4. The figure plots
anomalous baroclinic
pressure at the surface as a function of distance along the model
boundary
(abscissa) and time (ordinate). Distance along the boundary is
measured
counterclockwise from the southeast corner, labeled SE in the
plot; the northeast,
northwest, southwest, and southeast corners are labeled NE, NW,
SW, SE, respectively.
Clearly, the interdecadal oscillation is associated with a
disturbance that
propagates along the model boundary, counterclockwise around the
basin. Most of the
oscillation period is spent propagating along the weakly
stratified northern boundary,
and the northern parts of the eastern and western boundaries.
Propagation along the southern, and most of the eastern
boundary, is much more rapid. Clearly, in the case shown, the
disturbance propagates all
around the model boundary in one oscillation period. In section
3.3, we shall address
whether or not the disturbance must propagate all around the
model domain in order to maintain the oscillation.

It might be thought that the propagating disturbance evident in
Figure 4 is associated
with advection by the mean flow. This is not the case, as can
easily be understood by
noting that the mean flow is toward the east throughout the upper
part of the water
column north of $50^o$N, but is toward the west beneath. This is
not consistent with
the propagation along the northern boundary, which is westward
and almost in phase throughout the whole depth of the water
column.

Following \markcite{{\it Wajsowicz and Gill} [1986]}, and the
analysis given by
\markcite {{\it Winton} [1996]}, we interpret the boundary
propagating disturbance as a coarsely resolved, viscous,
baroclinic Kelvin wave.
\markcite{{\it Davey et al.}\ [1983]} discuss the properties of
these waves
in a continuous medium, and \markcite{{\it Hsieh et al.}\ [1983]}
discuss
the modifications when these waves propagate on a finite
difference grid, as in a
numerical model. The low-frequency form of these waves
described by \markcite {{\it Winton} [1996]} uses the Laplacian
mixing term
in the momentum equations to break the geostrophic balance, in
order to allow
divergence, and the associated vertical velocity,
necessary for wave propagation. This contrasts with inviscid
coastal Kelvin waves
[\markcite{{\it Gill}, 1982]} for which the local time derivative
terms are used
to break geostrophic balance and allow vertical motion. It should
also be noted that
whereas the velocity normal to the coast is zero for an inviscid
wave, this is not
the case for waves with viscosity.
\markcite {{\it Winton} [1996]} has noted that when the model
resolution is
insufficient to resolve the boundary layer, the waves take the
form of the
numerical boundary waves discussed by \markcite{{\it Killworth}
[1985]}.
Geostrophic balance is then broken by using the no-normal flow
condition
at the coast in the numerical computation of the divergence. An
example
is provided by the oscillation of \markcite{{\it Greatbatch and
Zhang} [1995]}. These
authors used the
planetary geostrophic model of \markcite{{\it Zhang et al.}\
[1992]}, which includes
explicit friction only in the vertically averaged part of the
flow. In
Greatbatch and Zhang's study, the vertically averaged flow is
zero, and
the momentum equations reduce to geostrophy. Implementation of
the no-normal flow
boundary condition breaks the geostrophic balance at the coast
and
allows propagation of the numerical boundary waves.
It is the existence of these waves that is responsible for the
oscillation in
\markcite{{\it Greatbatch and Zhang's} [1995]} paper.

An interesting aspect of Figure 4 is the special character of the
northeast corner
(marked NE). The almost complete elimination of the density
stratification along
the northern boundary strongly arrests wave propagation and gives
the
impression of disturbances being ``held up" in the northeast
corner. In this respect,
it is interesting to consider Experiment A0, which does not
oscillate. In the steady
state, there is a horizontal pressure gradient along the
eastern boundary associated with the surface eastward jet.
Westward propagation of
this jet is suppressed by the surface forcing, as demonstrated by
\markcite{{\it Winton} [1996]}, who shows an example where the
surface forcing is removed,
and a boundary wave immediately starts propagating along the
northern boundary.
We believe perturbations to the balance in the northeast corner
play a role in
initiating the wave propagation associated with the oscillation,
as we
demonstrate in subsection 3.3. Once
wave propagation is initiated, the slow propagation along the
northern boundary
leads to a considerable increase in the amplitude of the wave.
The slow propagation
is itself related to the very weak model stratification, and
plays a role in setting
the interdecadal timescale of the oscillation.

\begin{figure}[tbp]
\figurenum{5}
\figurewidth{20pc}
\begin{center}
\mbox{\epsfig{figure=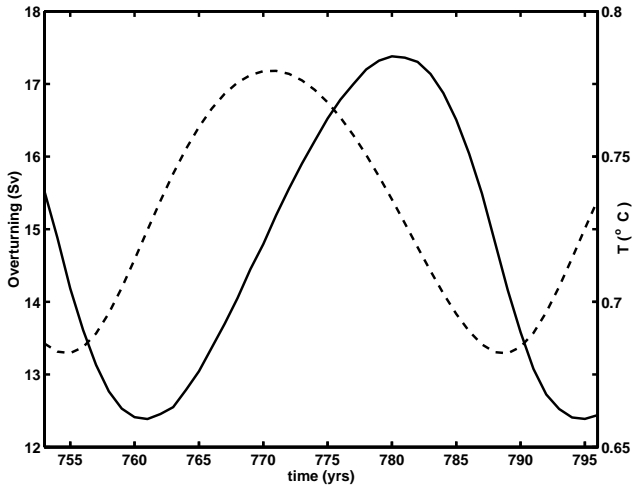,width=8.4cm}} 
\end{center}
\caption{Experiment A1. Time series of the maximum in the
overturning stream function (solid line),
and the temperature averaged over the region north of 35$^o$N
subtracted from the temperature averaged over the region south of
35$^o$N (dashed line).}
\end{figure}

In box models
[\markcite{{\it Stommel,} 1961;} \markcite{{\it Griffies and
Tziperman}, 1995}],
it is assumed that the strength of the thermohaline circulation
and the north/south density contrast vary in phase and are
proportional
to one another. By contrast,
a necessary consequence of the thermal wind relation is the
association of an
enhanced (reduced) east-west, rather than north-south, density
contrast with
enhanced (reduced) north-south flow
of the thermohaline circulation. In fact, the strength of the
thermohaline overturning and the north-south density contrast do
not vary in
phase in our model, as can be seen from Figure 5.
The solid line shows the time series of the maximum of the
overturning
stream function over one particular period of the oscillation.
The dashed line shows
the corresponding temperature averaged over the
northern part of the basin (north of $35^o$N), subtracted from
the
temperature averaged over the southern part
of the basin (south of $35^o$N). The choice of dividing latitude
is not important;
a similar result is obtained by using latitudes other than
$35^o$N, including dividing
latitudes that are close to the northern boundary.
The north/south temperature difference is a measure of
the north/south density contrast (recall that salinity is uniform
in this
experiment), with the maximum in the south/north
temperature difference corresponding to the maximum in the
north/south
density difference. It is obvious that the maximum in the density
contrast leads the
maximum in the overturning by about $90^o$. This is not
surprising in view of the
previous discussion. In particular, the maximum in the
overturning
occurs at a time when the ``east minus west" temperature
difference, and associated
baroclinic pressure gradient, reaches its peak along the
northern boundary (a consequence of thermal wind). The maximum in
this gradient
is associated with the propagation of a warm front along the
northern
boundary, at which time the high latitudes are already warmer
than in the mean.

The boundary adjustment process also has consequences for zonally
averaged models.
In these models, it is assumed that the east-west pressure
difference is
directly proportional to, and varies in phase with, the
north-south pressure gradient
[\markcite{{\it Wright and Stocker,} 1991}]. When the adjustment
process is active,
the assumption breaks down, as illustrated by the pressure fields
plotted in Figure 3.
We suggest that the ``in phase" relationship,
assumed in box models and zonally averaged models, is valid only
on timescales long compared to the adjustment timescale. For the
flat-bottomed ocean
models studied here, this timescale is decadal. Including
variable bottom topography
can alter the adjustment timescale because of the influence of
variable bottom topography
on the available wave modes, a topic of ongoing research. In a
similar way, the adjustment
timescale could be different in models of different resolution
[\markcite{{\it D{\"o}scher et al.}, 1994}], with advective
processes
playing a more important role at high resolution, as in the study
of
\markcite{{\it Gerdes and K{\"o}berle} [1995]}.

A question arises as to whether long, baroclinic
Rossby waves
play a role
in the adjustment process in our model experiments. In the study
of
\markcite{{\it Wajsowicz and Gill} [1986]}, these waves were
important on the
decadal timescale for spreading the influence of the eastern
boundary into the
ocean interior [\markcite{{\it Wajsowicz}, 1986}]. We have
carried out many experiments
using different model geometries in the hope of separating the
Rossby wave effect from
that of the boundary waves. These include basins with tilted
eastern and western boundaries, and also experiments using the
realistic coastline
of the North Atlantic. In all cases, any influence of Rossby
waves is secondary to
the influence of the viscous Kelvin wave propagating around the
boundary of the
model domain. Indeed, it is always the latter that dominates the
variation in the
thermohaline overturning in the model experiments. The lack of an
important role for
Rossby waves is consistent with \markcite{{\it Winton} [1996]},
who has showed that
interdecadal variability is also found in models run on an f
plane, for which there are
no Rossby waves.

\subsection{\bf 3.2. A Large-Amplitude Oscillation}

Experiment B0 in Table 2 is spun-up exactly as Experiment A0,
except that a smaller value ($1000$ m$^2$ s$^{-1}$) is used for
the
horizontal diffusivity. Following the spin-up, the surface
boundary condition is
switched to the diagnosed flux. Using the same model parameters
as the spin-up,
and initializing with the model state at the end of the spin-up,
an oscillation of a 30.5-year period develops. This shows that
zonal redistribution,
as discussed by
\markcite{{\it Cai et al.}\ [1995]}, is not always
necessary for the development
of oscillations. It is also clear that the model behavior is
sensitive to the
value of the horizontal diffusivity, consistent with the
sensitivity study of
\markcite{{\it Huang and Chou} [1994]}.

\begin{figure}[tbp]
\figurenum{6}
\figurewidth{20pc}
\begin{center}
\mbox{\epsfig{figure=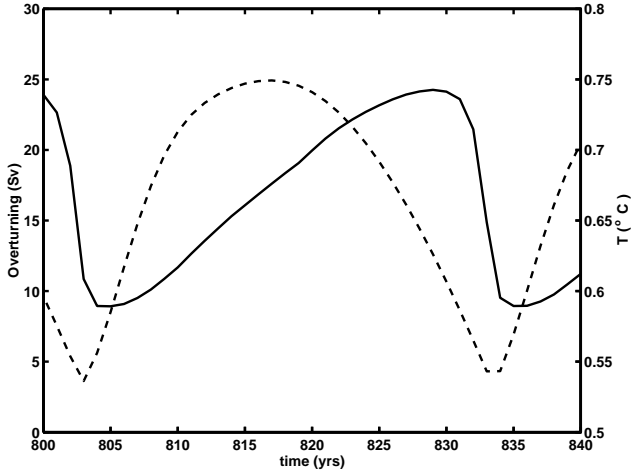,width=8.4cm}} 
\end{center}
\caption{Experiment B0. Time series of the maximum in the
overturning stream function (solid line),
and the temperature averaged over the region north of 35$^o$N
subtracted from the temperature averaged over the region south of
35$^o$N (dashed line).}
\end{figure}

The model variables in Experiment A1 undergo behavior that is
close to being
sinusoidal in nature (see Figure 5). In contrast,
the oscillation that develops in Experiment B0 has a distinctly
nonsinusoidal
behavior, an indication of strong nonlinearity. As can be seen in
Figure 6,
there is now a strong asymmetry between the gradual strengthening
phase of the
oscillation and its sudden collapse. The stronger nonlinearity
is also indicated by the much larger fraction of the mean
overturning that is taken by the oscillation amplitude (roughly
0.45 in
Experiment B0, compared to 0.17 in Experiment A1; see Table 2).
As in Experiment A1, the strength of the overturning lags the
north/south density contrast (see Figure 6), but the maximum in
the overturning is now
considerably delayed after the maximum in the north/south density
contrast and, in fact, occurs only a few years before the minimum
in that contrast.

As before, the oscillation is associated with the propagation
of a wave around the model boundary (see Figure 7). The gradual
strengthening of the overturning is associated with the slow
propagation of a
warm front along the northern boundary, the sudden collapse with
the rapid movement
of the front down the western boundary. Clearly, propagation of
the warm front along
the northern boundary (the strengthening phase) is quite
different from propagation
of a cold front. In Figure 7 the propagation of the warm front is
associated
with the steep gradient in baroclinic pressure that develops
along the
northern boundary around years 830 and 860. Close examination
shows that ahead of the
warm front,
the water is mixed to the bottom, indicating no stratification.
By contrast, behind the
front, convective mixing occurs over only part of the water
column. We believe it is the
strong contrast in the density stratification either side of the
front that gives the
wave its highly nonlinear character. (There is an analogy here
with a tidal river
bore associated with an incoming tide.) In particular, points
behind the wave front
have increased stratification and therefore increased (local)
gravity
wave speed, compared to points ahead of the front, resulting in a
steepening of the wave front.
\begin{figure}[tbp]
\figurenum{7}
\figurewidth{20pc}
\begin{center}
\mbox{\epsfig{figure=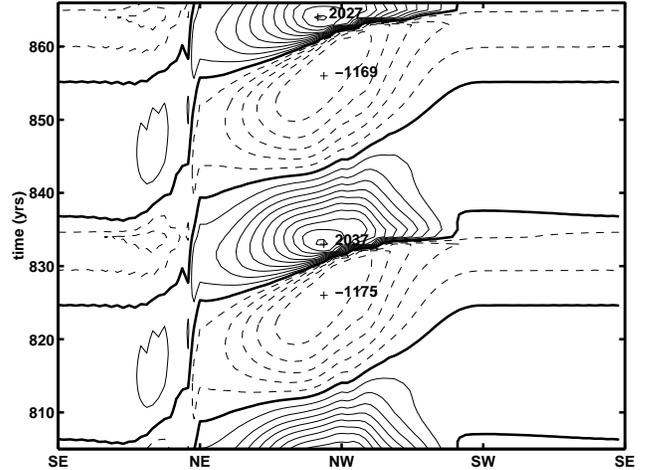,width=8.4cm}} 
\end{center}
\caption{Experiment B0. Contours of baroclinic pressure anomaly
(that is, difference from the mean averaged between the years
2586 to
2982) at level 1 (surface) as a function of distance along the
model boundary and time.  Distance is measured in a
counterclockwise direction from the southeast corner (marked SE).
The contour interval is 200 Pa.
Negative values are shown using dashed contours.
The maximum and minimum values are displayed (in Pa) for each
oscillation and identified with a cross.}
\end{figure}

\subsection{\bf 3.3. The Role of Wave Propagation Along Each
Model Boundary}

The oscillation in both Experiments A1 and B0 is associated
with the passage of a boundary wave once around the model domain
(see Figures 4 and
7). The question arises as to whether the wave must make a
complete
circuit of the boundary, and what role is played by the
boundaries in
maintaining the oscillations. For example, \markcite{{\it Winton}
[1996]} has suggested
that the oscillations are a consequence of thermal wind currents
impinging on
coasts with weak stratification. He points, in particular, to the
surface
eastward jet in the northern part of the basin impinging on the
eastern boundary as
being the source of decadal variability in models that oscillate
under
constant surface flux
boundary conditions. To address these issues, we have carried out
four
experiments in which the wave propagation is suppressed along
each of the
southern, eastern, northern, and western boundaries. These
experiments
are labeled
Experiments B0a,b,c,d, respectively, in Table 2. In the case of
the southern boundary
(Experiment B0a in Table 2), the experiment is identical
to Experiment B0, except that along the four rows of grid points
nearest the southern
boundary, the temperature field is relaxed back to its initial
value (that is, the state
at the end of the spin-up). The timescale for this relaxation is
1.5 days nearest the
boundary and 6.7 days along the outermost row of grid points. In
the case of the other
boundaries, the temperature field is relaxed back to its initial
value only along
the two rows of grid points nearest the boundary, the timescale
for the relaxation being 2.5 days.

We begin with Experiment B0a, in which wave propagation is
suppressed along the
southern boundary. We find that an oscillation
still occurs. The character of the oscillation is very much like
that in Experiment B0,
except that the amplitude is increased, and the period is now
37.2 years instead of
30.5 years. The fact that the period and amplitude are increased
shows that
allowing the wave to make a complete circuit of the model domain
does have an effect,
but it is not crucial for the existence of the oscillation. It
is, nonetheless,
of interest that preventing wave propagation along the remote,
tropical boundary of the
model domain does have an influence on high-latitude variability
in the model.

When wave propagation is suppressed along the northern boundary
(Experiment B0c),
there is no oscillation, as we expect. The interesting
experiments are
B0b and B0d. In the former, wave propagation is suppressed along
the
eastern boundary. The oscillation still occurs and has a period
of 36.2 years and
amplitude of 10.3 Sv, both of which are similar to the period and
amplitude in
the case with wave propagation suppressed along the southern
boundary.
In B0d, wave propagation is suppressed along the western
boundary. This time
no oscillation is found, a result that is surprising in view of
\markcite{{\it Winton's} [1996]} suggestion that the oscillations
are maintained by the
eastward flowing jet impinging on the eastern boundary. If the
oscillations
were being generated on the eastern boundary, we should expect to
see a wave propagating
along the northern boundary from its eastern end, even though the
wave propagation is
being suppressed on the western boundary. It is also interesting
that the oscillation
is still found in Experiment B0b, even though the temperature
field is being maintained
at a constant value along the eastern boundary. We conclude that
it is the western
boundary that is the most important for maintaining the
oscillation.

Propagation of the wave southward along the western boundary
perturbs the western boundary current. We suggest that the
perturbed western boundary
current generates an anomaly in temperature that is
then advected across the basin and initiates the wave propagation
at the northeast corner.
For example, at year 775 in Figure 3a, a surface low-pressure
anomaly is propagating
southwards down the  western boundary. The northward
intensification of the western
boundary
current leads to the warm, high-surface pressure, anomaly to the
south, which
in turn,
links to the eastern boundary as the warm front associated with
the strengthening phase
of the oscillation starts propagating along the northern
boundary.
The special character of the northeast corner was noted in
subsection 3.1, where we
commented
that in a steady state, the surface forcing must exactly balance
the tendency for wave
propagation along the northern boundary. It follows that a
disturbance to this balance
can lead to propagation.  Amplification of the disturbance then
takes place
as the wave makes its way slowly along the weakly stratified
northern boundary.
In this way, we see the link to Winton's original suggestion that
it is the eastern
boundary that is important.
In the case with propagation suppressed along the eastern
boundary, the wave starts
to develop from a point on the northern boundary just east of the
northeast corner,
which is consistent with the above idea. Allowing the wave to
propagate from the
western boundary, along the southern boundary and on up the
eastern boundary, as in
Experiment B0, adds an additional perturbation to the northeast
corner region,
allowing the northern boundary propagation to start slightly
earlier. This effect is
demonstrated by comparison with Experiments B0a and B0b, in which
the propagation is suppressed on the southern and
eastern boundaries. Both Experiments B0a and B0b give
oscillations
of similar period and amplitude but with a longer period than in
Experiment B0.
It is also interesting to note that the above
interpretation shows the relationship between oscillations under
constant flux, and
oscillations under mixed surface boundary conditions to be
discussed next. In the latter,
eastward movement across the basin interior is a
pronounced feature of the oscillation. \markcite{{\it Weaver and
Sarachik} [1991]} have
associated this eastward movement with an advective process.

\section{\bf 4. Mixed Boundary Conditions}

\begin{figure}[tbp]
\figurenum{8}
\figurewidth{20pc}
\begin{center}
\mbox{\epsfig{figure=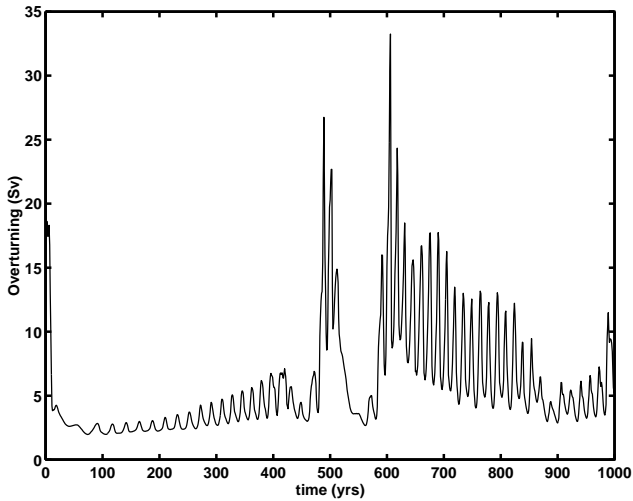,width=8.4cm}} 
\end{center}
\caption{Experiment C1. Time series of the maximum in the
overturning stream function in the case run under mixed surface
boundary conditions.}
\end{figure}

We now consider what happens on a switch to mixed boundary
conditions (Experiment C1
in Table 2).
In other words, at the end of the spin-up, the surface boundary
condition
on salinity is replaced by the diagnosed flux, but the restoring
boundary
condition is maintained on temperature. Figure 8 shows a
time series of the maximum in the overturning stream function.
Upon the
switch of boundary conditions, the circulation undergoes a rapid
collapse
(a polar halocline catastrophe; [\markcite{{\it Bryan}, 1986]}).
After roughly 500 years, there is a flush, followed by subsequent
collapses
and flushes. Leading up to the first flush, there is a slow
recovery of the overturning circulation, with decadal
oscillations
superposed. These oscillations have the same character as those
discussed by
\markcite{{\it Weaver and Sarachik} [1991]},
as can easily be verified, and have a period that decreases from
26 years
initially, to about 17 years. The oscillations are associated
with large changes
in the surface heat flux. The changes in surface heat flux act to
keep the surface temperature close to the
restoring temperature given by equation (2), as required by the
restoring
boundary condition. As such, oscillations under mixed boundary
conditions are associated
with surface buoyancy flux forcing that varies interdecadally, in
contrast to
the oscillations discussed in section 3, for which the surface
buoyancy flux
is constant in time.

\begin{figure*}
\figurenum{9}
\figurewidth{35pc}
\begin{center}
\mbox{\epsfig{figure=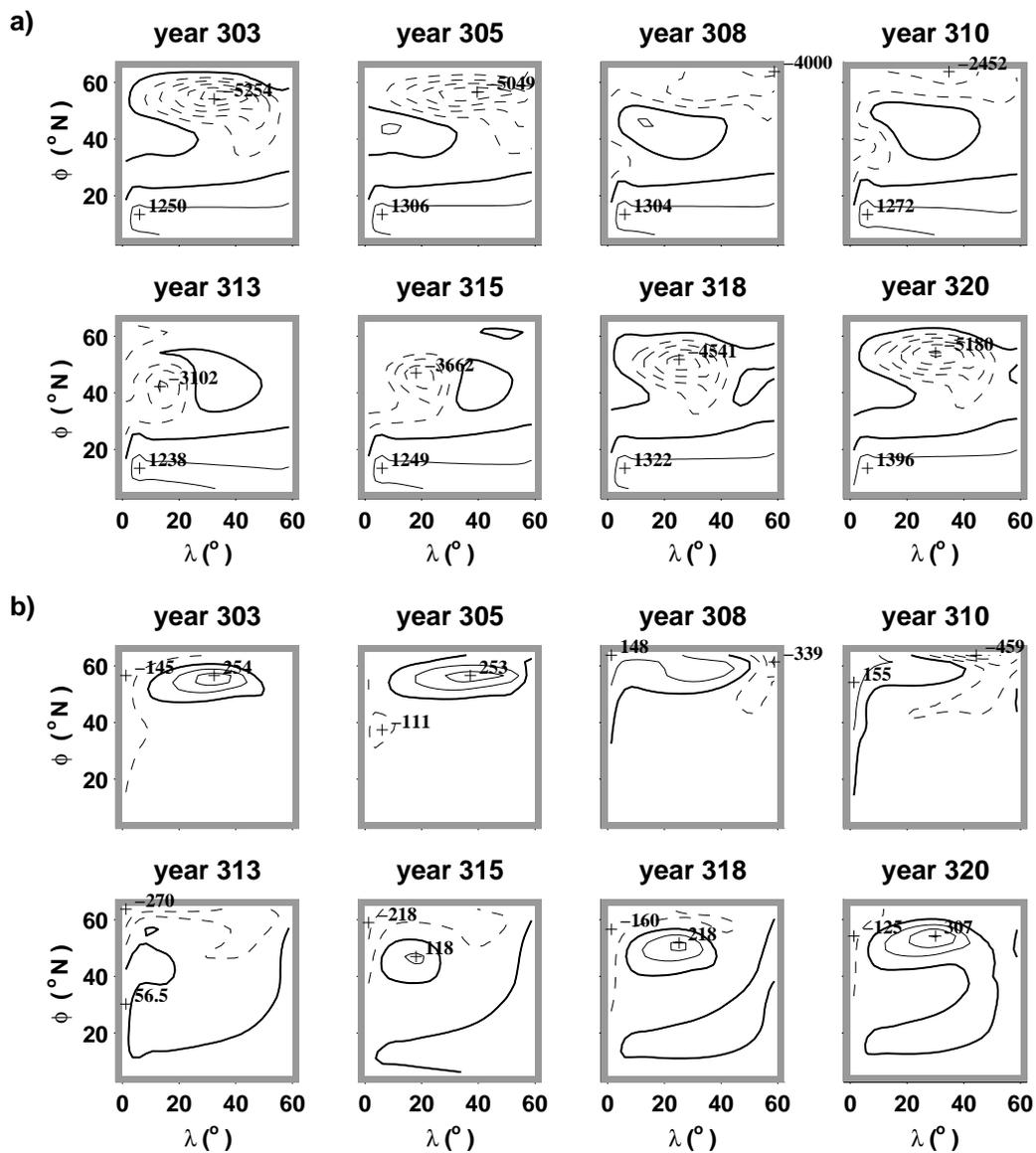,width=13.8cm}} 
\end{center}
\caption{Experiment C1. Snapshots of total baroclinic pressure
for (a) level 1 (0 m to 46 m deep) with contour interval of 1000
Pa and (b) level 11 (1607 m to 2070 m deep) with contour interval
of 100 Pa. Negative values are shown using dashed contours.
The maximum and minimum values are displayed on the contours (in
Pa) and identified with a cross.}
\end{figure*}

\begin{figure*}
\figurenum{10}
\figurewidth{35pc}
\begin{center}
\mbox{\epsfig{figure=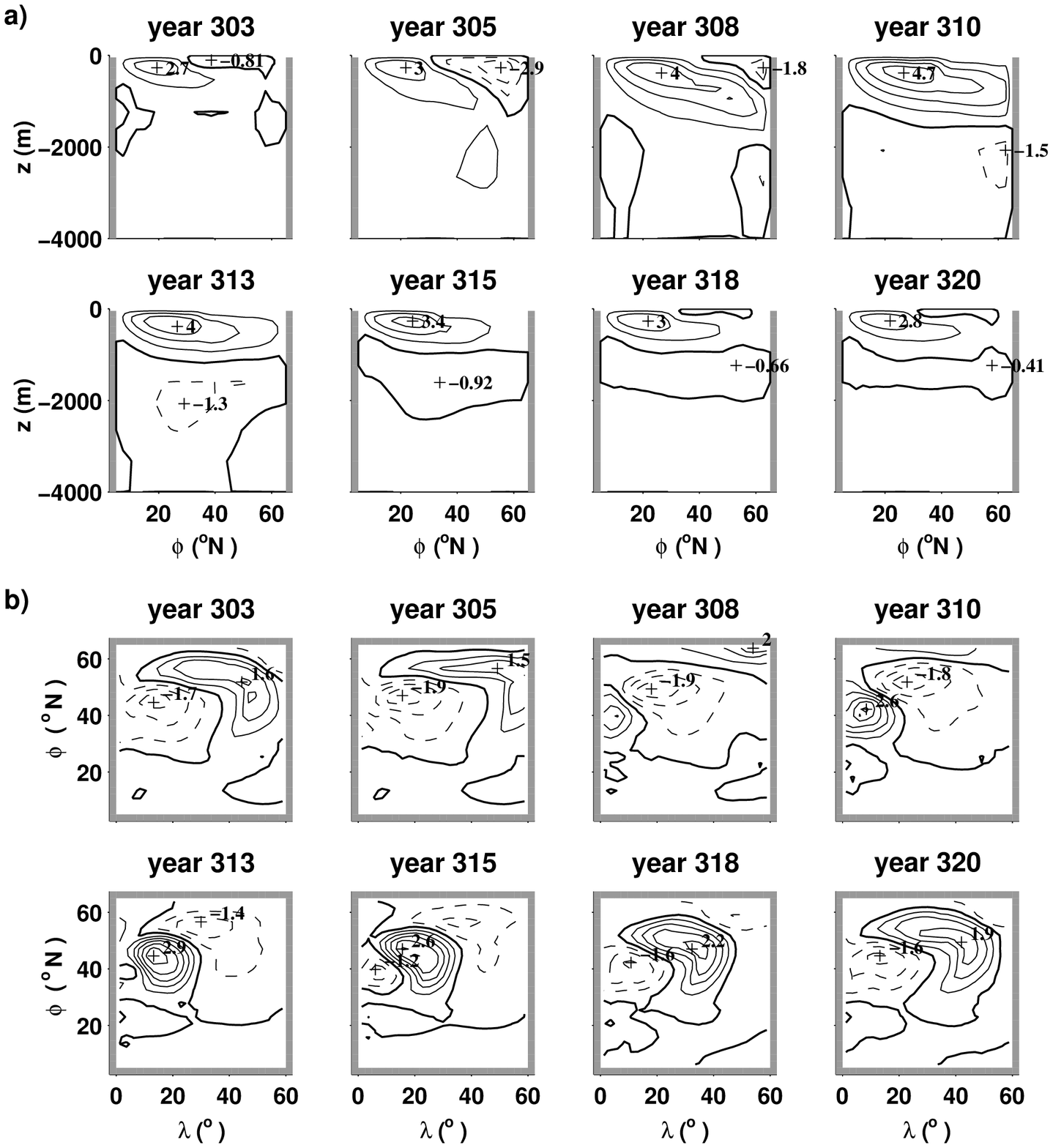,width=13.8cm}}  
\end{center}
\caption{Experiment C1. Snapshots of (a) the overturning stream
function with contour interval 1.0 Sv and (b) the surface
salinity anomalies with contour interval $0.5^o/_{oo}$.
The meridional overturning has its minimum and maximum strength
at years 303 and 310, respectively.
Anomalies are differences from the mean state
calculated by averaging from years 303 to 338.
Negative values are shown using dashed contours.
The maximum and minimum values are displayed on the contours (a)
in Sv and (b) in $^o/_{oo}$ and identified with a cross.}
\end{figure*}

As before, the baroclinic pressure proves to be
a useful diagnostic. Figure 9 shows snapshots of the baroclinic
pressure
(the time mean has not been removed) at the surface and at level
11 (1838.5-m depth).
The large features in midbasin are associated with
salinity-dominated density
anomalies that dominate
the northern part of the basin, move eastward across the basin at
midlatitudes,
and then westward along the northern boundary (compare Figure 9
with Figure 10b).
In midbasin, where the amplitude is
large, the vertical structure of these features is like that of
the first baroclinic mode,
an indication of which is the different sign of the pressure
anomalies at level 11
compared to the surface. Along the boundaries, the vertical
structure is more
complex, sometimes showing structure like that of higher
baroclinic modes.
Evidence of this is the multicell structure of the thermohaline
overturning
plotted in Figure 10a.
Figure 11 plots the anomaly in the
baroclinic pressure as a function of distance around the model
boundary (as in
Figure 4) at both the surface and at level 11 (1838.5-m depth).
Once again, we see
propagation in a counterclockwise direction around the model
domain.
This is particularly evident at level 11, where the oscillation
is associated with the
propagation of a disturbance all around the model domain,
including the
southern boundary (it should be noted that the location of the
zero contour in Figure 11b is complicated by trends in the model
variables
associated with the gradual build up to the flush).
Indeed, at depth, the characteristics of the oscillations under
mixed boundary conditions have many similarities with those under
constant surface flux
boundary conditions.

\begin{figure}[tbp]
\figurenum{11}
\figurewidth{20pc}
\begin{center}
\mbox{\epsfig{figure=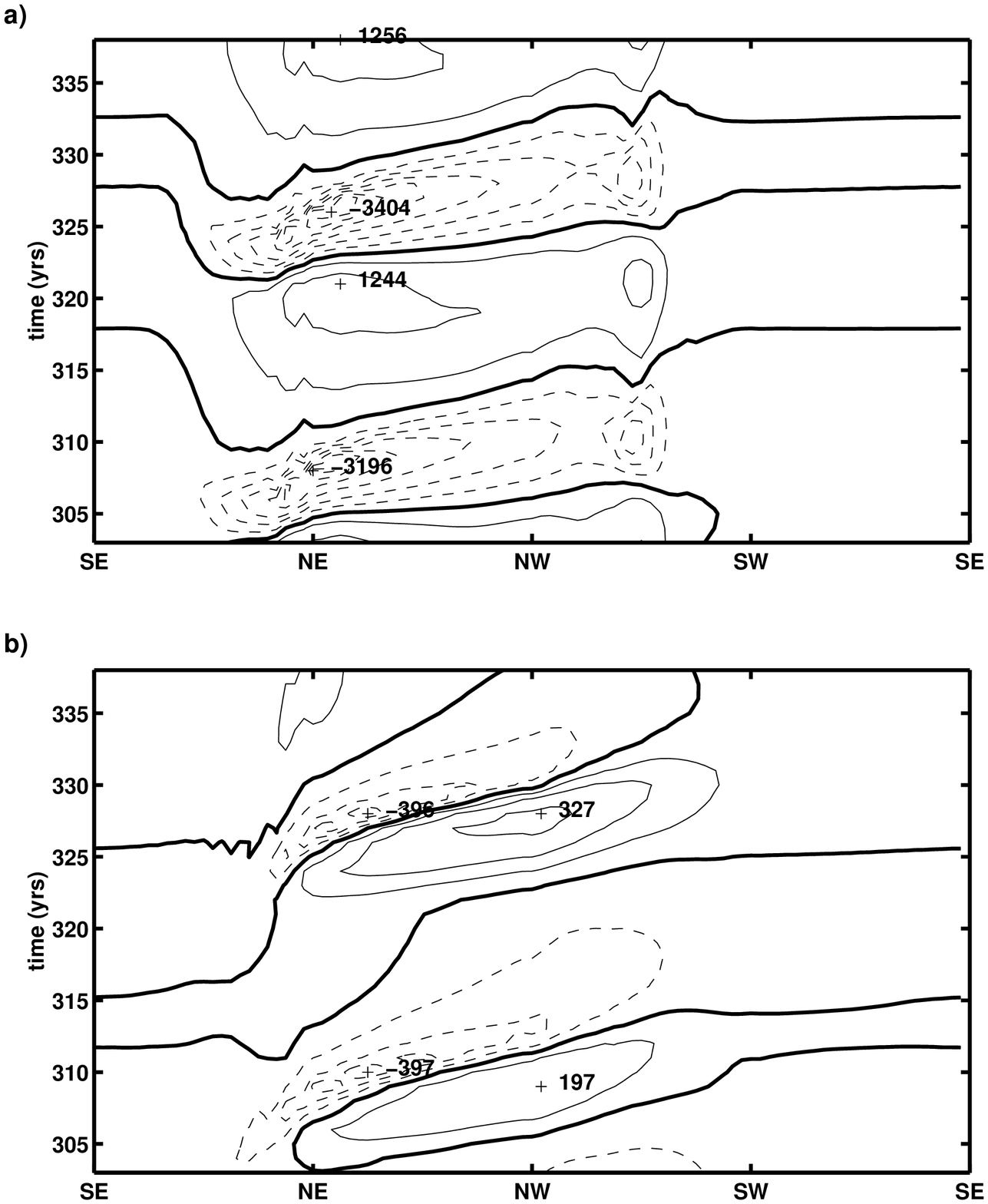,width=8.4cm}} 
\end{center}
\caption{Experiment C1. Contours of baroclinic pressure anomaly
as a function of distance along the model boundary and time at
(a) level 1 (0~m to 46~m deep), with contour interval 500 Pa,
and (b) level 11 (1607~m to 2070~m deep), with contour interval
100 Pa. Distance is measured in a counterclockwise direction from
the southeast corner (marked SE).
Negative values are shown using dashed contours.
The maximum and minimum values are displayed (in Pa) for each
oscillation and identified with a cross.}
\end{figure}
We contend that viscous Kelvin wave propagation plays an
important role in the westward
movement of the anomalies across the northern boundary. We point
first to the evidence
of propagation along the boundary noted in the previous
paragraph.
The manner and characteristics of the propagation are consistent
with the
viscous Kelvin wave propagation noted in section 3.
Furthermore,
the similarity at depth between the
structure of the mixed boundary condition oscillations discussed
here and the constant
flux oscillations discussed in section 3 adds further weight to
the argument,
particularly since the deeper depths are more isolated from the
influence of the
changing surface heat flux associated with the restoring boundary
condition.
Also, given the predominantly geostrophic nature
of the flow in the model, the pressure fields plotted in Figure 9
clearly indicate
regions of convergence and divergence that propagate along the
model boundaries, and
are the signature of wave propagation. We suggest that the
convergence/divergence
at the boundary is associated with the vertical movement of the
anomalies that forms
part of the explanation for the oscillation put forward by
\markcite{{\it Weaver and Sarachik} [1991]}.

An interesting aspect of the midbasin, eastward propagating
anomalies is the association of strong surface heat loss with the
positive salinity
anomalies. The regions of strong surface heat loss move eastward
with the
positive salinity anomalies. Freshwater input at high latitudes,
associated with
the surface freshwater flux boundary condition, is mixed downward
by convective
overturning due to surface heat loss. Reduced surface heat loss
allows the
freshwater to accumulate at the surface, leading to a negative
salinity anomaly
[\markcite{{\it Zhang et al.}, 1993}]. Similarly, strong surface
heat loss is associated
with strong vertical mixing, which in turn, mixes surface
freshwater downward,
leading to a positive salinity anomaly at the surface. Viewed in
this way,
the eastward moving salinity anomalies can be regarded as
being forced by the surface heat flux anomalies (that is, this
part of the
oscillation can be regarded as the forced part), whereas the
westward movement along
the northern boundary, which we have associated with viscous
Kelvin wave
propagation, can be regarded as the internal, unforced part of
the oscillation.
Of course, this argument
does not address why the midbasin anomalies in surface heat flux
move eastward.
To understand this, it is necessary to consider the combined
ocean-atmosphere system
implied by the mixed surface boundary conditions. Certainly,
existence of a surface
freshwater anomaly would lead to reduced surface heat loss over
the anomaly
by the same mechanism by which a polar halocline catastrophe
occurs
[\markcite{{\it Zhang et al.}, 1993}]. Similarly, a positive
salinity anomaly is
associated with enhanced surface heat loss at the surface, since
the surface water
is warmer due to deep convective mixing. We suspect the salinity
anomalies are advected eastwards, as suggested by
\markcite{{\it Weaver and Sarachik} [1991]} and carry the surface
heat flux
anomalies with them.  However, because the anomalies completely
dominate the total
pressure field north of $30^o$N (see Figure 9), the details of
the advective mechanism
are not trivial, a point we shall return to in a later
manuscript.

Figure 12 compares the strength
of the overturning circulation with the north/south density
contrast. Once again,
we see that the two quantities do not vary in phase, the
thermohaline overturning lagging the north-south density
difference. Given the large anomalies that are found
north of $35^o$N (Figure 9), it is not surprising that this time
there is some
dependence on the choice of dividing latitude used for the
calculation.
We feel, however, that the choice of $35^o$N is probably the most
appropriate, since then the anomalies associated with the
oscillation occur almost
entirely north of the dividing latitude, the mean density south
of the dividing
latitude changing little during the oscillation. With this choice
of dividing latitude,
the thermohaline overturning lags the north-south density
contrast by almost $180^o$.

\begin{figure}[tbp]
\figurenum{12}
\figurewidth{20pc}
\begin{center}
\mbox{\epsfig{figure=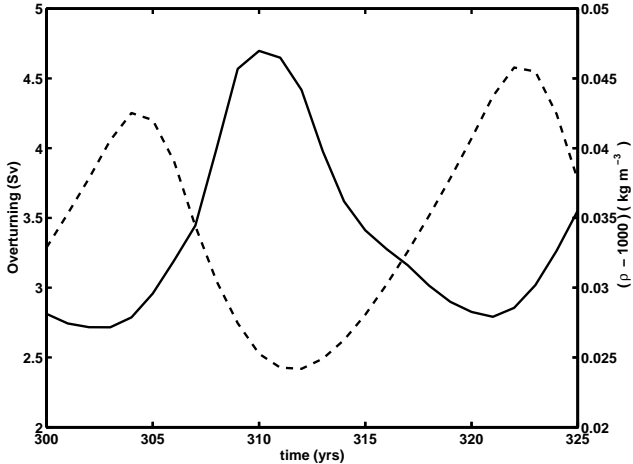,width=8.4cm}}  
\end{center}
\caption{Experiment C1. Time series of the maximum in the
overturning stream function (solid line), and the density
averaged over the region north of 35$^o$N subtracted from
the density averaged over the region south of 35$^o$N (dashed
line).}
\end{figure}

We can understand the phase relationship by comparing Figure 12
with
the baroclinic pressure fields plotted in Figure 9. At years 303
and 305, the thermohaline
overturning is at its minimum, and in fact, there is a shallow
reverse cell
north of $40^o$N (see Figure 10a). At this time, a large,
positive density
anomaly, associated with saline water, occupies the northern half
of the basin
(Figure 10b). This is associated with a large cyclonic eddy at
the surface, with
anticyclonic circulation at level 11 (Figure 9). The presence of
the cyclonic eddy,
and its associated positive density anomaly, explains why the
north-south density
contrast is enhanced, even though the thermohaline overturning is
weak.
It should also be noted that the deep convective mixing is
strong,
in association with the dense water of the cyclonic eddy, even
though the thermohaline
overturning is a minimum (this particular out-of-phase
relationship is a feature of the
ice-ocean oscillation of \markcite{{\it Zhang et al.}\ [1995]},
which also uses mixed
surface boundary conditions). The weak, surface-confined reverse
cell of the thermohaline
circulation is associated with the impingement of the cyclonic
eddy on the eastern
boundary, implying that the surface pressure is lower on the
eastern than on the
western boundary. It should be noted that most of the circulation
associated with the
cyclonic eddy makes no contribution to the thermohaline
overturning, with the
surface northward flow in the east being cancelled by surface
southward flow in the
west (it is the east-west pressure difference between the
boundaries that matters).
The increase in the strength of the thermohaline circulation is
associated with the
passage of a boundary wave along the weakly stratified part of
the eastern
boundary and the northern boundary. Pressure decreases at level
11 and increases
at the surface with the passage of the wave, the surface lagging
level 11
by several years.
At year 310, the
thermohaline overturning reaches its maximum. The northward
surface flow toward
the northern boundary, with southward flow away from the northern
boundary at level 11,
can be seen in Figure 9. At this time, a reduced density anomaly,
associated with
relatively fresh water, is present in the basin interior (Figure
10b) and explains the
minimum in the north-south density contrast.

\section{\bf 5. Summary and Conclusions}

The oceanic thermohaline circulation transports
roughly half the
heat from low to high
latitudes required to maintain the Earth's radiation balance
[\markcite{{\it Gill}, 1982}].
For the north-south flow to be in geostrophic balance, there must
be an east-west
pressure difference across the ocean basin. A fundamental
question is how the
east-west pressure difference is established, given an initially
imposed north-south
density gradient. This problem has been studied by
\markcite{{\it Wajsowicz and Gill} [1986]} and, more
recently, by
\markcite{{\it Winton} [1996]} and is closely related to the
adjustment problem
considered by
\markcite{{\it Davey} [1983]}. In flat-bottomed ocean models the
first stage of the
adjustment is carried out by viscous baroclinic Kelvin waves
[\markcite{{\it Wajsowicz and Gill}, 1986]}.
The second stage involves internal Rossby waves
\markcite{{\it Wajsowicz} [1986]}. \markcite{{\it Wajsowicz and
Gill} [1986]}, use a Kelvin wave adjustment timescale of only
months, much less than the
interdecadal timescale. On the other hand, the initial density
field
used by \markcite{{\it Wajsowicz and Gill} [1986]} had vertical
density stratification
at all latitudes,
facilitating wave propagation all around the model domain. A
feature of the models we
have considered [see also
\markcite{{\it Winton}, 1996}) is that the high-latitude density
stratification
is weak or nonexistent, greatly impeding internal Kelvin wave
propagation. In fact,
the presence of weak stratification, due to deep convective
mixing, so impedes wave
propagation that the adjustment by Kelvin waves now takes place
on an interdecadal timescale. We have seen evidence of viscous
Kelvin wave propagation in model experiments
run under constant surface heat flux (salinity kept uniform and
constant) and under
mixed surface boundary conditions (a strong restoring boundary
condition on the
surface temperature, and a constant flux boundary condition on
the surface salinity).
We suggest that oscillations under constant surface flux are
self-sustained by
perturbations to the western boundary current arising from the
southward
propagating boundary wave along the western boundary. These
perturbations are then
advected to the northeast corner and play a role in reinitiating
the wave propagation.
Under mixed surface boundary conditions, salinity-dominated
density anomalies move
eastward across the interior of the basin and then westward along
the northern
boundary. We suggest the latter is associated with viscous Kelvin
wave propagation.
Under mixed surface boundary conditions, the westward advective
phase is amplified by
the changing surface heat flux in response to the surface
salinity anomalies.
Under constant surface flux, the amplification of the oscillation
is associated
with the wave propagation along the weakly stratified northern
boundary.

We have not addressed the question of why oscillations are found
under some surface
flux fields but not under others [\markcite{{\it Cai et al.},
1995]}.
Experiment B0 is an example of an oscillation that occurs
when the surface restoring boundary condition used for the
spin-up is replaced
by the diagnosed surface flux, all other aspects of the model
being the same as
in the spin-up. On the other hand, Experiment A1 is an example
where it was necessary to
zonally redistribute the diagosed flux to obtain oscillations.
The two cases differ only
in the value of the horizontal diffusivity (see Table 2). We
suggest the horizontal
diffusivity is an important parameter for determining whether
self-sustained
oscillations occur
or not. The choice of horizontal diffusivity is, in turn,
dependent on the model
resolution and the choice of other mixing parameterizations,
indicating that these
choices can also influence the ability to obtain self-sustained
oscillations in a model.
Detailed discussion of this point is beyond the scope of the
present paper.

Throughout this paper, we have argued that boundary wave
propagation is essential for
the existence of the variability we have described.
This is very different from a situation in which variability is
generated by a mechanism independent of boundary waves. In such a
case,
boundary waves could be excited as forced waves, but would then
be symptoms of the variability, rather than essential to its
existence.
\markcite{{\it Rahmstorf et al.}\ [1996]}
have drawn an instructive anology between constant surface flux
oscillations in
three-dimensional models,
and the thermal ``loop" oscillator of \markcite{{\it Welander}
[1967]}. The loop
oscillator consists of the closed loop of fluid in the vertical
plane, cooled
at the top and heated at the bottom. A cold anomaly
in the sinking branch will accelerate the flow. The anomaly then
passes quickly
through the heating region, but only slowly through the cooling
region, maintaining
the oscillation. The crucial ingredient is the time delay between
the strength of the
flow and the density contrast between the top and bottom of the
loop. In the case of the
loop oscillator, the time delay is provided by the fluid inertia.
In three-dimensional
ocean models, the time delay is provided by the boundary wave
propagation associated
with the thermohaline adjustment process, as indicated by the out
of phase relationship
between the strength of the overturning circulation and the
tropical-polar density
contrast (Figures 5, 6, and 12).

The out-of-phase relationship demonstrated in Figures 5, 6, and
12 stands in
sharp contrast to the assumption in box models
[\markcite{{\it Stommel,} 1961;} \markcite{{\it Griffies and
Tziperman}, 1995}],
that the thermohaline circulation and tropical-polar density
contrast are in phase and
proportional to one another. The adjustment process also
invalidates the assumption
in zonally-averaged models that the east-west pressure difference
is directly
proportional to the north-south pressure gradient
[\markcite{{\it Wright and Stocker,} 1991}].
We believe both assumptions are valid only on
timescales long compared to the adjustment timescale. For the
flat-bottomed
ocean models we have considered, the adjustment timescale is
clearly
decadal. Including variable bottom topography may change the
adjustment timescale
because of the influence of the bottom topography on the
available wave modes, and is
a topic of ongoing research.

Although we have concentrated on the role of viscous baroclinic
Kelvin waves,
the studies by
\markcite{{\it D{\"o}scher et al.}\ [1994]} and
\markcite{{\it Gerdes and K{\"o}berle} [1995]}
point to the importance of advection, rather than boundary waves,
in setting the
interdecadal timescale of the adjustment in higher resolution
models (although
the problems studied in these papers do not involve wave
propagation along a
weakly stratified, high-latitude boundary). In fact,
it is clear that a major failing of course resolution models is
the inadequate way
in which they represent the coastal wave guide. This applies not
only to
the way in which wave processes are represented, but also to the
representation of shelf/slope currents, such as the Labrador
Current and the
Deep Western Boundary Undercurrent.
Given the importance of the wave guide
demonstrated in this paper, and that of \markcite{{\it Winton}
[1996]}, it is clear that
studies are required to test the robustness of interdecadal
variability in
models to both increasing resolution, and a more realistic
representation of coastal,
shelf/slope, processes.
Recent data studies (\markcite{{\it G. Reverdin et al.}},
Decadal variability of hydrography in the upper northern North
Atlantic 1948-1990, submitted to {\it Journal of
Geophysical Research}, 1996)
point to the importance of shelf/slope currents as a source
region for
interdecadal variability observed in the North Atlantic, again
pointing to the need
for more realistic
representation of shelf/slope regions in models.

\acknowledgments

Funding to R.J.G from NSERC, NSERC/WOCE, and AES in the form of a
Science Subvention award,
and a grant from the Canadian Institute for Climate Studies, are
acknowledged.
Comments from Stephen Griffies and two anonymous reviewers led to
improvements in the manuscript.

\end{document}